# DEUTERON FROZEN SPIN POLARIZED TARGET FOR nd EXPERIMENTS AT THE VdG ACCELERATOR OF CHARLES UNIVERSITY

N.S. Borisov[1], N.A. Bazhanov[1], A.A. Belyaev[4], J. Brož[2], J. Černý[2], Z. Doležal[2],
A.N. Fedorov[1], G.M. Gurevich[3], M.P. Ivanov[1], P. Kodyš[2], P. Kubík[2], E.S. Kuzmin[1],
A.B. Lazarev[1], F. Lehar[5,6], O.O. Lukhanin[4], V.N. Matafonov[†1], A.B. Neganov[1],
I.L. Pisarev[1], J. Švejda[2], S.N. Shilov[1], Yu.A. Usov[1] and I. Wilhelm[2]

[1]*Joint Institute for Nuclear Research, Dubna, Russia;*
[2]*Institute of Particle and Nuclear Physics, Faculty of Mathematics and Physics, Charles University in Prague; V Holešovičkách 2, CZ18000 Prague 8, Czech Republic;*
[3]*Institute for Nuclear Research, Russian Academy of Sciences, Moscow, Russia;*
[4]*NSC KIPT, Kharkov,*
[5]*SPP DAPNIA CEA Saclay, F-91190 Gif sur Yvette, France*
[6]*IEAP CTU, Horska 3a/22 CZ 12800 Prague 2, Czech Republic*

*Abstract.* A frozen spin polarized deuteron target cooled by the $^3$He/$^4$He dilution refrigerator is described. Fully deuterated 1,2-propanediol was used as a target material. Deuteron vector polarization about 40% was obtained for the target in the shape of a cylinder of 2 cm diameter and 6 cm length. The target is intended for a study of 3N interactions at the polarized neutron beam generated by the Van de Graaff accelerator at the Charles University in Prague.

## 1. Introduction

The Charles University proton polarized target (PPT) [1] has been used previously to measure spin–dependent total cross section differences $\Delta\sigma_T$ and $\Delta\sigma_L$ in neutron-proton scattering at the 16.2 MeV polarized neutron beam of the VdG accelerator (Fig.1) [2, 3].

---

[†] Deceased.

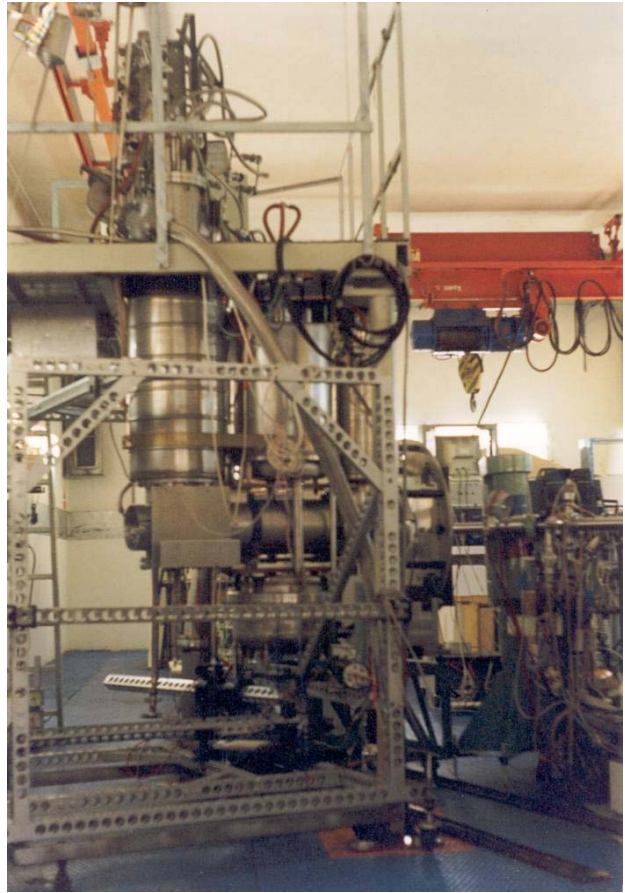

Fig.1. The frozen polarized target in the beam area

Recently, a new project has been started studying three-nucleon interactions in the final state [4] with the use of 16 MeV polarized neutron beam scattering off the polarized deuteron target (PDT).

To realize this new project a target with frozen polarization of deuterons has been developed using $^3$He/$^4$He dilution refrigerator of the existing PPT. This work required a significant improvement of refrigerator characteristics as well as thorough choice of the target material and modification of the equipment to measure deuteron polarization.

## 2. General description

PDT is a facility consisting of a $^3$He/$^4$He dilution refrigerator, $^3$He and $^4$He pumping system, a superconducting magnetic system and PC controlled equipment to build-up and measure the target polarization. A general view of the target is shown in Fig. 1.

The refrigerator with a horizontal tail containing the target material has a module design (Fig. 2) including three autonomous matching units: a $^3$He pre-cooling, condensing and pumping unit (1), a cryostat (2), and a dilution unit (3). The units are interconnected through vacuum seals, thus providing easy access to every unit for separate tests and repair. The tail of the refrigerator is aligned along the neutron beam axis. In details the refrigerator design is described in Section 3.

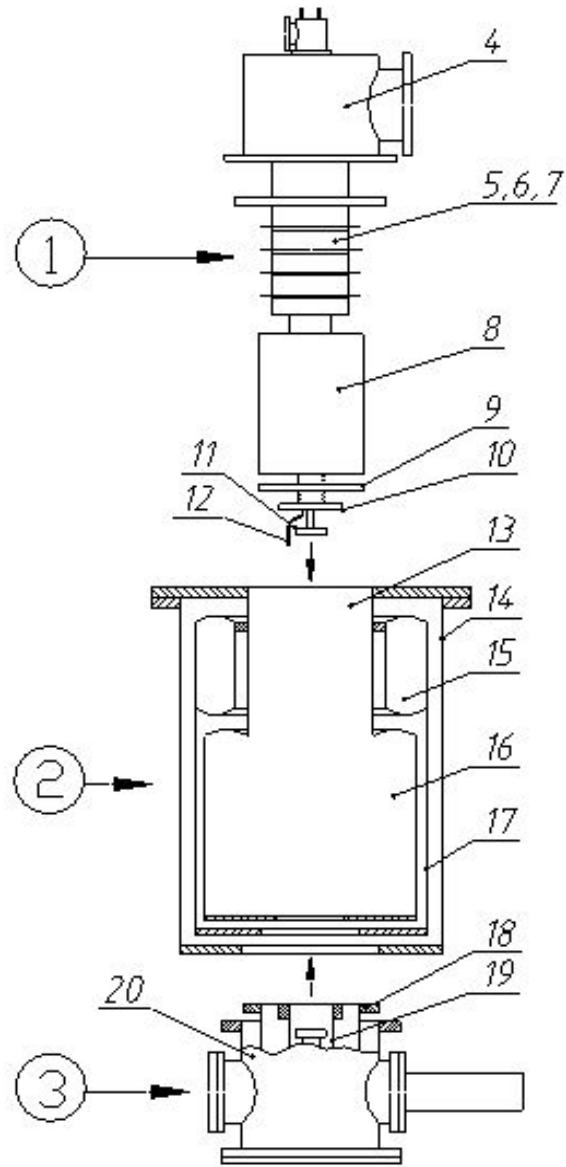

Fig. 2. Three main units of the dilution refrigerator

The $^3$He pumping system of the refrigerator generates $^3$He circulation in the dynamic polarization mode (~$10^{-2}$ mole/s) and in the frozen spin mode (~ $2\times10^{-3}$ mole/s) and consists of roots pumps WS-2000 and WS-250 (Leybold), and a rotational pump H-2030 (Alcatel). $^4$He from the volume of the condensing (1 K) bath is pumped by two pumps WS-500 and H-2060.

The target magnetic system (Fig. 3) consists of two superconducting magnets: a movable solenoid with a horizontal field and a large aperture stationary dipole magnet with a vertical field direction. The solenoid which is being put on the tail of the refrigerator in the dynamic polarization mode generates the magnetic field of 2.7 T in the target volume with homogeneity of about $10^{-4}$.

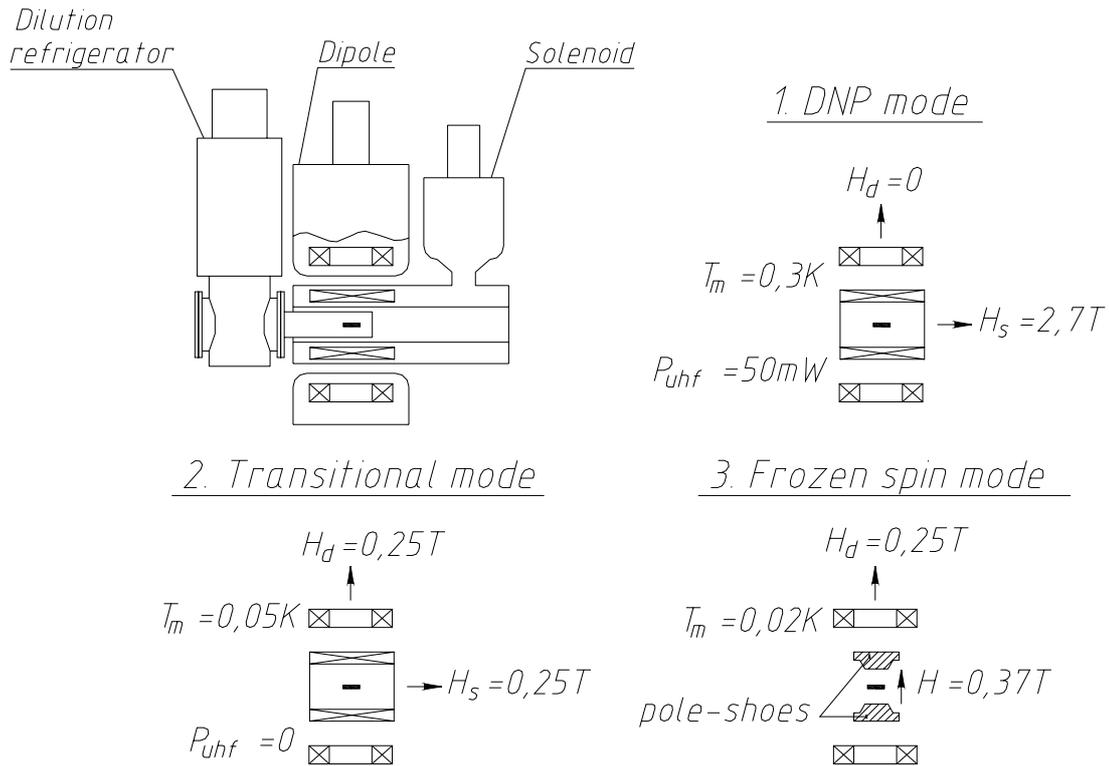

Fig. 3. Magnetic system of the target and magnetic field configurations in different working modes

The superconducting dipole is used as a holding magnet in the frozen spin mode. Its cryostat consists of two parts connected by two tubes. This design provides ±50° aperture in the vertical plane and nearly 360° aperture in the horizontal plane, which is essential for certain types of experiments. The dipole magnet (with an addition of two Armco iron pole-shoes) generates a 0.37 T magnetic field in the target volume with homogeneity of about 1%.

Combinations of magnetic fields being used in different working modes are illustrated in Fig. 3. If the longitudinal deuteron polarization is required then after the microwave pumping of polarization in the DNP mode (Fig. 3.1) the target is being cooled below 50 mK in the frozen spin mode and left in a horizontal solenoid field of 0.5 – 2.7 T. If, however, the transverse target polarization is used then after the DNP mode the solenoid field is decreased to about 0.25 T and the field of the holding magnet is increased up to the same magnitude (Fig. 3.2). After that the solenoid field is decreased to zero, and the cryostat of the solenoid is moved horizontally along the beam axis and further in a transverse direction to remove it from a vicinity of the target. An insertion of the pole-shoes permits to increase the vertical holding field to approximately 0.37 T, which provides an elongated nuclear relaxation time.

### 3. Dilution refrigerator

The refrigerator design is illustrated in Figs. 2 and 4. The $^3$He pre-cooling, condensing and pumping unit (1) consists of a cap (4) connectible to external pipelines, $^3$He and $^4$He pumping tubes (5, 6) with gas heat exchangers (7), a $^3$He condensing unit (8), flanges (9, 10) for connecting to the cryostat and dilution unit, respectively, and is completed by a still cover (11) and a $^3$He capillary (12).

The 1 K bath (Fig. 4., 21) is filled with $^4$He through the vapour-cooled coil pipe (41) and adjustable needle valve (24). $^3$He gas cooled in the upper heat exchanger comes to a coil pipe (42) placed inside the $^3$He pumping tube, condenses in a sintered copper condenser (22) and comes to the dilution unit through the adjustable needle valve (23) and restrictor (40).

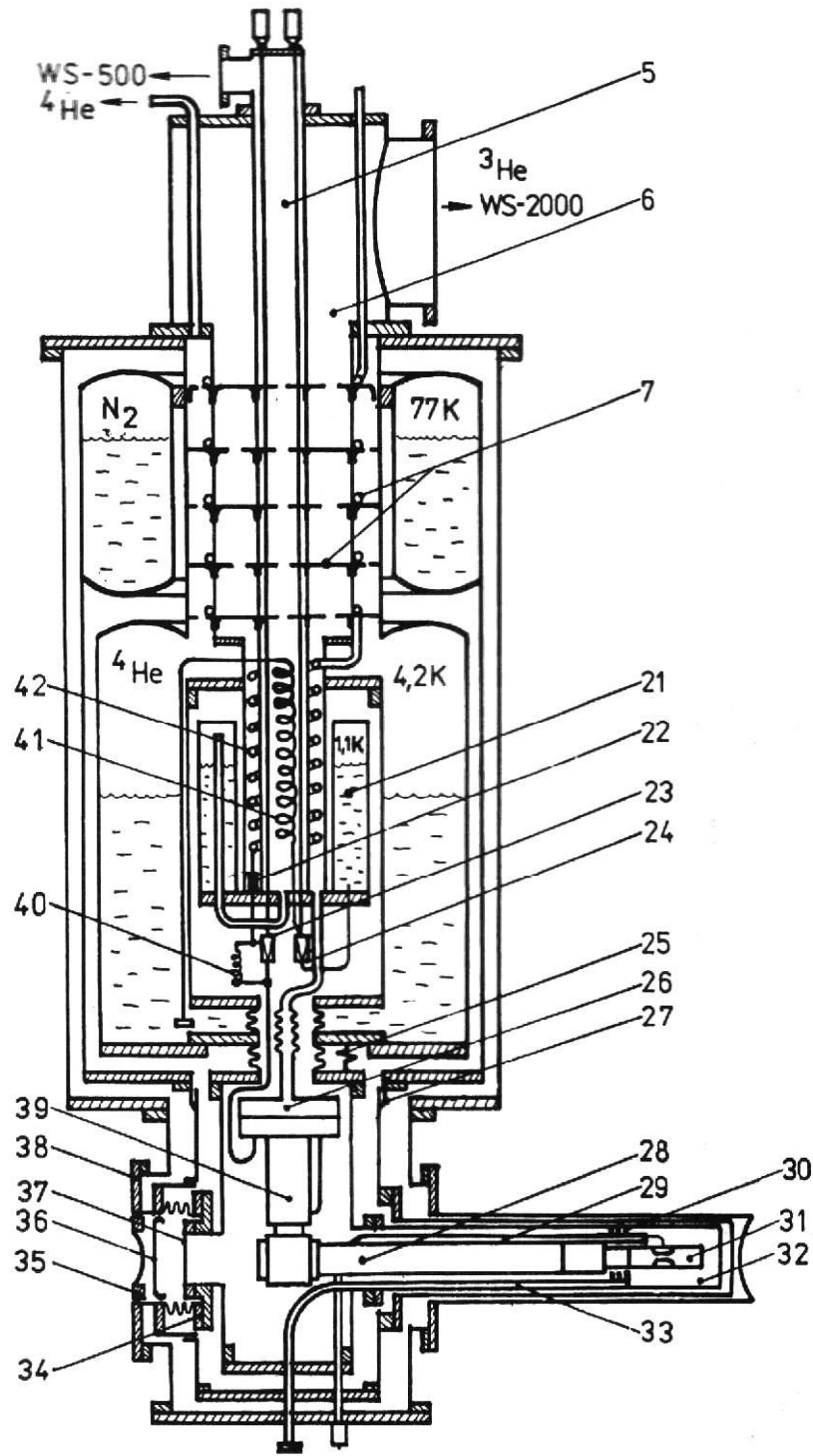

Fig. 4. Schematic view of the vertical cryostat of the dilution refrigerator

The cryostat is provided with a 17 l liquid nitrogen vessel (15) and a 19 l liquid $^4$He vessel (16) ensuring 30 hours of continuous work without refilling. Holes provided in the lower parts of the helium vessel, the liquid nitrogen cooled shield (17) and the outer vacuum jacket permit to interconnect corresponding elements of the pumping and dilution units.

The dilution unit (Figs. 2, 4, and 5) has a jacket (20), nitrogen (18) and helium (19) shields, a $^3$He still (26), heat exchangers (28) and (39) and a mixing chamber (31), rigidly connected by tension wires and posts. Sample loading is performed through a lock chamber (38). A microwave cavity formed by a tail portion of the helium shield and a microwave choke (30) is fed through a wave guide (33). A RF-circuit coil is connected to the NMR equipment by a cable (29).

Thermal contacts of the nitrogen and helium shields are made of flexible copper wire bundles (25, 27). The wave guide and the NMR cable are cooled through the point thermal contacts at 80 K, 4 K and 0.8 K.

The lock chamber (38) is used to load and extract a target sample with a melting point of about -50°C. During the sample loading operation a flange (34) connected to bellows is sealed in with the helium shield, and after cooling the cryostat to the liquid nitrogen temperature the lock chamber channel and a space inside the helium shield are filled with helium gas at atmospheric pressure. Then covers (35, 36, 37) are removed using a special device, which limits air inflow to the cooled elements. After sample loading the covers are reinstalled and helium gas is pumped out. Then the flange (34) can be removed from the helium shield, thus breaking the thermal coupling through the lock chamber channel and merging the chamber volume with that of the cryostat vacuum jacket.

Internal arrangement of the dilution unit is illustrated in more detail in Fig. 5. This unit consists of three easily disconnectible parts. First of them includes a still body (45), a preliminary heat exchanger (39) and a connecting assembly (57). The second part consists of a main heat exchanger (28) and a mixing chamber (31). The third part intended for sample loading and providing a $^3$He/$^4$He solution channel has a shape of long hollow barrel (56) of 30 mm diameter with a Teflon ampoule screwed on its end (53) containing the target material. Introduction of the ampoule into the environment cooled down to the liquid nitrogen temperature is performed through the lock chamber using a special device, which permits to seal an indium gasket in vacuum.

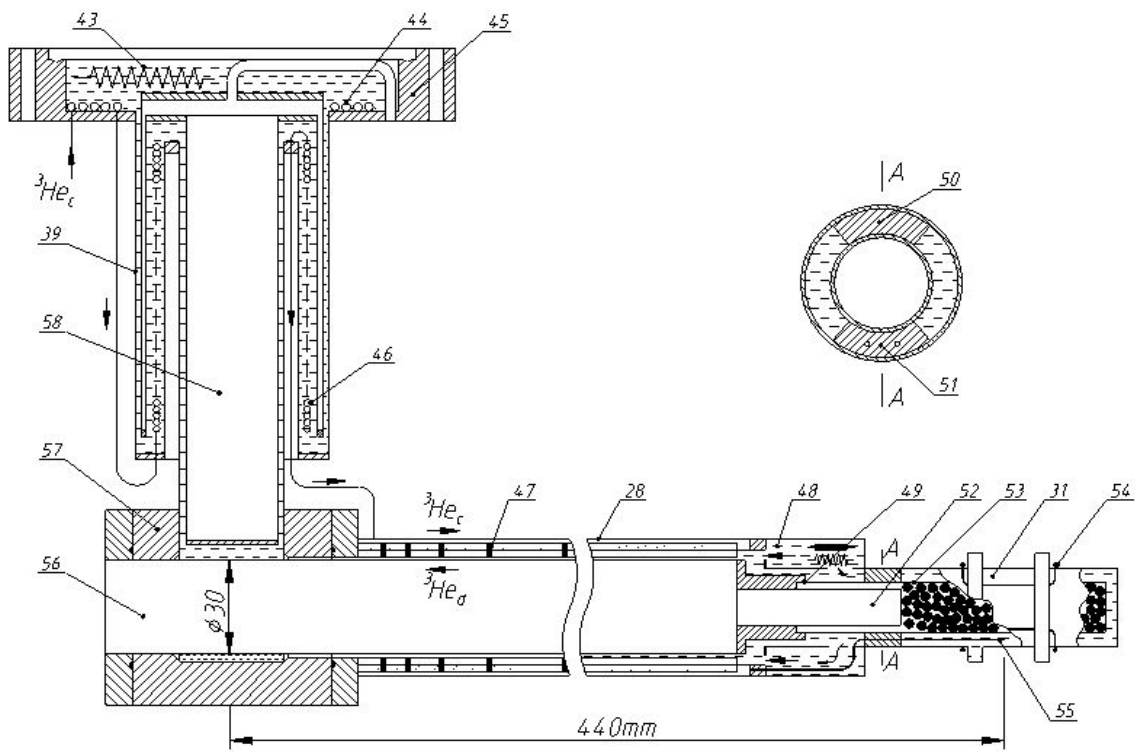

Fig. 5. Low temperature part of the dilution refrigerator

During the connection of the dilution step and the pumping unit the $^3$He still is formed, when the cover seals the still body (45) through the indium gasket, providing a plane box of 110 mm diameter and about 100 cm$^3$ volume. A heater (43) in a shape of a flat spiral of constantan wire is placed at approximately half of the still height. To monitor the still temperature a 10 Ω Speer carbon resistor placed on the heater body is used. The tube heat exchanger (44) cooling the returning $^3$He flow is mounted on the still bottom. It has a volume about 150 cm$^3$ and geometric impedance of $2.8 \times 10^8$ cm$^{-3}$.

The preliminary heat exchanger contacting the still bottom has a 60 mm external diameter and a 150 mm long body, a hollow insert (58) and a double helix coil pipe for $^3$He flow. The total volume of this heat exchanger equals 750 cm$^3$ and the impedance of the $^3$He channel is $2 \times 10^8$ cm$^{-3}$. This design has to provide most effective work in qualitatively different regimes corresponding to high and low $^3$He circulation rates.

Connecting assembly (57) has a massive stainless steel body welded to the preliminary heat exchanger and being connected through indium seals to the main heat exchanger (28) and the barrel (56).

The main heat exchanger (28) prepared of a sintered copper powder with ~20 μm diameter grains has a calculated active surface of about 4 m$^2$. An annular channel for the $^3$He/$^4$He solution between the walls of the heat exchanger and the barrel (56) has a width of ~0.5 mm. The cooled $^3$He flows along the outer annular channel of about 0.1 mm width.

Copper powder is sintered on both outer and inner surfaces of a thin walled copper-nickel tube of 36 mm diameter. These surfaces are preliminary galvanicly covered by a thin copper layer. The sintering was performed in hydrogen atmosphere at 820°C in a stainless steel matrix. The heat exchanger is sectioned by piles of thin stainless steel foils (47), which were inserted during a filling of the matrix by the copper powder. This technique provides thermally isolated sections due to a large number of consecutive Kapitza resistances formed between the foils.

To make the heat exchanger work more effective, its total area is distributed unevenly between the sections. As calculations show, even if a small number of sections is used, it is possible (by using the optimum area distribution) to make the efficiency of the step heat exchanger close to an ideal continuous heat exchanger. In our case the section area increases approximately 15 times from the beginning to the end of the heat exchanger.

An arrangement of the mixing chamber and NMR measuring cell is shown in Fig. 4. $^3$He cooled in the heat exchangers is fed through two 1.5 mm diameter capillaries (55) under a perforated 20 mm diameter ampoule (53) containing the target material. A lower annular sector (51) in the mixing chamber is intended for fastening the $^3$He capillaries while an upper sector provides a stock of $^3$He in the upper part of the mixing chamber. An RF-circuit coil is placed outside the mixing chamber Teflon body (31). $^3$He flows through the side annular channels (see AA cross section in Fig. 5) and then through windows, covered by metal mesh against microwave irradiation, to the measuring cell (48) where carbon resistor thermometers and a heater are placed, and further to the heat exchanger channel. A bottom of a spout of a hollow cylinder (52), upper and lower annular sectors, and a microwave choke (Fig. 4) form together the back wall of the microwave cavity. A portion of the microwave power penetrates the side $^3$He/$^4$He channels as far as the ampoule threaded plug (49), and also through the annular clearance between the microwave choke and the mixing chamber (Fig. 4).

### 4. Temperature measurements

For a normal target operation a continuous temperature monitoring of various target units is necessary. A special device providing reliable temperature measurements for the cryogenic facility disposed inside the accelerator hall has been developed. A scheme of the device is shown in Fig. 6.

Calibrated carbon resistors (Allen Bradley, Speer and TVO) were used as temperature sensors. Sensor resistance measurements were performed using an alternate measuring current. A frequency of the measuring current (25 Hz) was obtained by dividing the main frequency (50 Hz) which helped to effectively suppress the power-supply noise. To exclude the sensor overheating by the measuring current the sensor voltage has been kept below 100 μV for all resistance ranges.

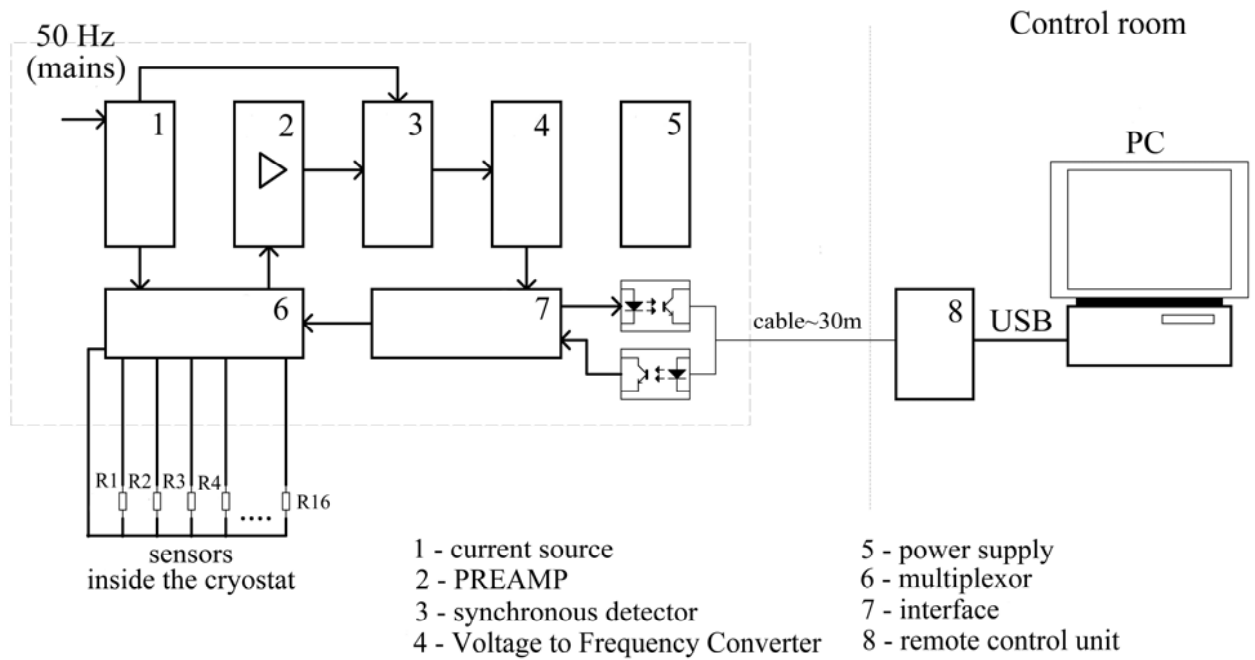

Fig. 6. Block diagram of the temperature measurement system.

The device consists of two units: a measuring unit placed immediately on the target cryostat to minimize the length of connecting wires to thermometers, and a control unit placed in the accelerator control room and connected to a personal computer. A distance between the units might be as large as 100 m. The connecting cable has an optical decoupling to prevent a ground loop noise.

The measuring unit contains a 16 input commutator, 4 position range switch on reed relays, a low-noise pre-amplifier, a measuring current generator, a synchronous detector with integrator, a voltage-to-frequency converter, a control circuit with interface, and a power supply (220V/50Hz).

The control unit contains a microcontroller, an interface with a measurement circuit, and USB interface for the connection to the personal computer. This unit is powered through the USB cable.

The computer program permits to control the temperature of the cryostat at 16 points for practically unlimited period of time (the duration of work is limited by the capacity of the PC hard disk only).

## 5. Deuteron polarization build-up and measurement

*5.1. Microwave pumping of polarization*

A system providing the microwave pumping of deuteron polarization consists of a microwave generator, a wave guide inside and outside the dilution refrigerator, and a multimode cavity containing the target material. The microwave cavity is formed by the walls of the 1 K copper shield into which the neusilber wave guide is introduced. To suppress a microwave power leakage to the heat exchangers, groove shaped wave restrictors are used. Outside the refrigerator the wave guide includes an attenuator to adjust the microwave power introduced to the cavity and a directional coupler with a detector head to measure frequency and power of the microwave oscillations. The microwave generator is a 4 mm wavelength oscillator using an ATT diode placed inside the Invar cavity; with an output power of above 100 mW. A frequency tuning in 73.0 - 75.5 GHz range is provided by the cavity piston. Additional smooth tuning in the range up to 30 MHz can be obtained by the change of the voltage at the varactor diode. An availability of the electronic adjustment permits to tune the frequency of the microwave generator more accurately in accordance with pumping maxima for both signs of the polarization and to use a frequency modulation of the generator output power (the modulation frequency 0 – 100 kHz, the deviation up to 5 MHz). The frequency modulation of the microwave power is necessary to obtain higher deuterium polarization in the deuterated 1,2-propanediol used as a target material.

*5.2. Calculation of deuteron polarization value*

A polarization value for particles with the spin $I$=1 can be extracted from the measured intensities of transitions between states with spin projections 1, 0, -1 (+1↔0, 0↔-1) using the formula $P = (r^2-1)/(r^2+r+1)$ where $r$ is a ratio of the transition intensities. To obtain the transition intensities from the DMR absorption spectrum, in which the lines corresponding to different transitions are partially overlapped, the procedure described in [5] was used. A fit of the DMR spectrum by the function describing the shape of the line was performed with the use of ROOT software (CERN). In our case a spectrum was decomposed into 5 lines. Two lines correspond to transitions (+1↔0, 0↔-1) for C-D couplings in 1,2-propanediol molecules, two other lines represent analogous transitions for O-D couplings. Additionally, a dispersion correction was calculated using Kramers-Kronig relation, which was introduced with some coefficient. The value of this coefficient was determined together with the ratio of the transition intensities as a result of the fit. The existence of a small (<3%) dispersion was an evidence of the correct Q-meter tuning. A total error of the polarization measurement does not exceed 3-4%. In Fig. 7 a typical DMR spectrum together with the fit is shown.

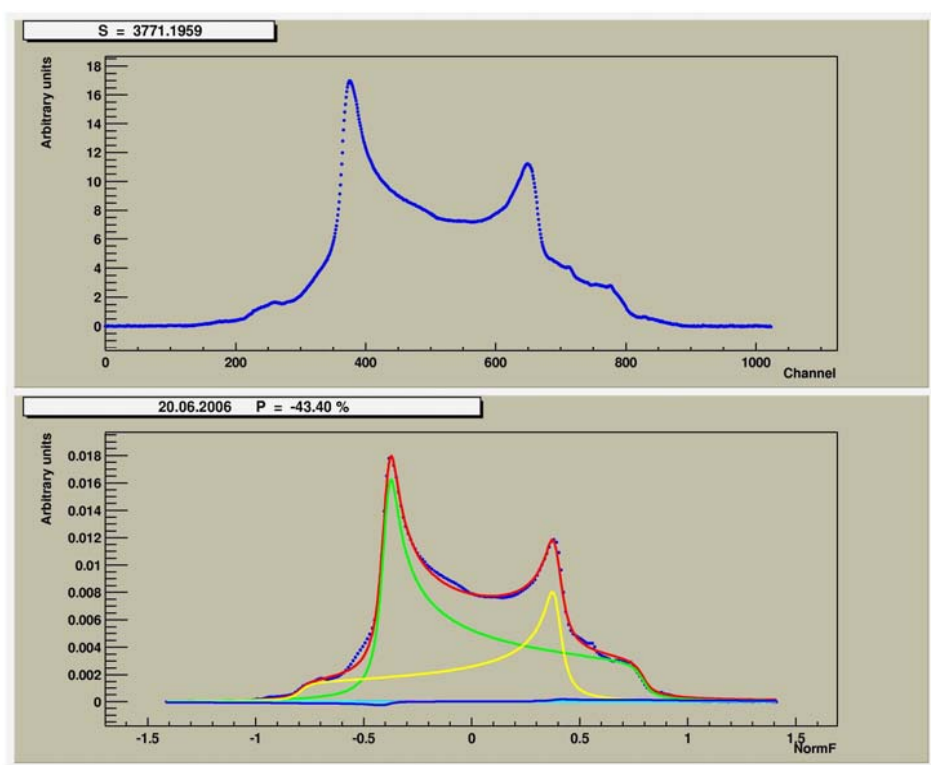

Fig. 7. Typical DMR spectrum (upper part) and fit results (lower part)

## 6. Conclusion

The target with a frozen deuteron polarization has successfully started its operation at the polarized neutron beam of the Charles University (Prague) accelerator. The target facility includes three helium cryostats: the dilution refrigerator and two superconducting magnets providing longitudinal and transverse deuteron polarization. Deuterated 1,2-propanediol with a paramagnetic Cr (V) impurity having a spin concentration about $10^{20}$ cm$^{-3}$ is used as a target material. The maximum deuteron vector polarization achieved exceeds 40%. Experiments studying the three-nucleon interaction contribution are in progress.

## 7. Acknowledgements

The authors are expressing their gratitude to Profs. A. Sissakian, V.Kadyshevsky, S.Gerasimov, A. Olshevski, A. Kovalík and P. Sorokin for their help and support of the work. The authors are also grateful to V. Kolomiets, O. Shcevelev and L. Doležal for their help during the installation of the facility.
This work has been partially supported by the research plan of the Ministry of Education, Youth and Sports of the Czech Republic under the Reg. No. MSM0021620859.